\documentclass[a4paper,10pt]{scrartcl}

\usepackage{latexsym}
\usepackage{amsmath}
\usepackage{amsfonts}
\usepackage{amssymb}
\usepackage{graphicx}
\usepackage[english]{babel}
\usepackage{wasysym}
\usepackage{units}
\newcommand{\p}{\partial}
\newcommand{\reff}[1]{(\ref{#1})}
\newcommand{\vs}[1]{\vspace{#1mm}}
\newcommand{\mt}[1]{\mathcal{#1}}
\newcommand{\vsO}{\vspace{.1cm}\hfill\\}
\newcommand{\vsT}{\vspace{.2cm}\hfill\\}
\newcommand{\q}{\quad}
\newcommand{\qq}{\qquad}
\newcommand{\ph}{\phantom1}

\newcommand{\sss}{\scriptscriptstyle{(\omega)}}
\newcommand{\primo}{{\scriptscriptstyle\prime}}

\title{\Large  A NOVEL APPROACH TO\\ LORENTZ GAUGE THEORY}

\author{Nakia Carlevaro$^{\;a,\;b}$ and Giovanni Montani$^{\;b,\;c,\;d,\;e}$\vsT
\emph{\footnotesize $^a$Department of Physics, Polo Scientifico -- Universit\`a degli Studi di Firenze,}\vs{-2.5}\\
\emph{\footnotesize INFN -- Section of Florence, Via G. Sansone, 1 (50019), Sesto Fiorentino (FI), Italy}\\
\emph{\footnotesize $^b$ICRA -- International Center for Relativistic Astrophysics,}\vs{-2.5}\\
\emph{\footnotesize c/o Dep. of Physics - ``Sapienza'' Universit\`a di Roma}\\
\emph{\footnotesize $^c$ Department of Physics - ``Sapienza'' Universit\`a di Roma, Piazza A. Moro, 5 (00185), Rome, Italy}\\
\emph{\footnotesize $^d$ENEA -- C.R. Frascati (Department F.P.N.), Via Enrico Fermi, 45 (00044), Frascati (Rome), Italy}\\
\emph{\footnotesize $^{e}$ ICRANet -- C. C. Pescara, Piazzale della Repubblica, 10 (65100), Pescara, Italy}\vsO
{\footnotesize\ttfamily nakia.carlevaro@icra.it\quad montani@icra.it}
}
\date{}
\begin{document}
\maketitle

%
\hrule
\begin{abstract} \textbf{Abstract:}
This paper is devoted to introduce a gauge theory of the Lorentz Group based on the analysis of isometric diffeomorphism-induced Lorentz transformations. The behaviors under local transformations of fermion fields and spin connections (assumed to be coordinate vectors) are analyzed and the role of the torsion field in a curved space-time is discussed.
\end{abstract}
\hrule

\vspace{1cm}
\section{General Remarks}
In this work, we will demonstrate that an isometric diffeomorphism can \emph{formally} induce a local Lorentz rotation, thus standard spin connections $\omega_\mu^{\ph ab}$ have no longer a gauge role in this framework (being only a function of tetrads and behaving like vectors under the diffeomorphism-induced Lorentz rotation). New gauge connections $A_\mu^{\ph ab}$ have to be introduced into the dynamics to appropriately recover the Lorentz invariance of the scheme, when spinor fields are taken into account. Furthermore, in the First-Order Approach, the geometrical identification of the Lorentz Group (LG) gauge fields with a suitable bein projection of the contortion field is allowed when a non-standard interaction term between generalized connections and these gauge fields is postulated, if fermion matter is absent. 

{\footnotesize\textbf{\emph{Notation:}} Greek indices (\emph{e.g.}, $\mu=0,1,2,3$) change as tensor ones under general coordinate transformations (\emph{i.e.}, world transformations); Latin indices (\emph{e.g.}, $a=0,1,2,3$) are the tetradic indices and refer to Lorentz transformations.}\vspace{-1mm}

\section{Review on the Tetradic Formalism and Spin Connections} Here we want to review the tetradic approach to General Relativity (GR). In such a scheme, the metric $g_{\mu\nu}$ and the tetrads $e^{\ph a}_{\mu}$ are linked by the usual relation $g_{\mu\nu}=\eta_{ab}\,e_{\mu}^{\ph a}\,e_{\nu}^{\ph b}$, where $\eta_{ab}$ is the local Minkowski metric. Projecting tensor fields from the 4-dimensional manifold to the Minkowskian space-time allows us to emphasize the local Lorentz invariance of the scheme in presence of spinor fields. In fact, fermions transform like a particular representation $S$ of the LG, \emph{i.e.}, $\psi\to S\psi$,
\begin{align}\label{LG}
S=I-\tfrac{i}{4}\;\epsilon^{{a}{b}}\,\Sigma_{{a}{b}}\;,\qq\qq
\Sigma_{ab}=\tfrac{i}{2}\,[\gamma_{a},\gamma_{b}]\;,\qq\qq
[\Sigma_{cd},\Sigma_{ef}]=i\mathcal{F}^{ab}_{cdef}\,\Sigma_{ab}\;,
\end{align}
where the $\Sigma_{ab}$'s and the $\mathcal{F}^{ab}_{cdef}$'s are the generators and the structure constants of the LG, respectively and $\epsilon^a_b(x)$ is the infinitesimal Lorentz rotational parameter. To assure the Lorentz covariance of the spin derivative $\p_\mu\,\psi$, connections $\omega_{\mu}^{\ph ab}$ must be introduced to define a covariant derivative as 
\begin{equation}\label{spin_connections}
D^{\sss}_\mu=\p_\mu+\Gamma^{\sss}_\mu\;,\qquad
\Gamma^{\sss}_\mu=\tfrac{1}{2}\;\omega_{\mu}^{\ph ab}\, \Sigma_{ab}\;,\qquad
\omega_{\mu}^{\ph ab}=e^{a\nu}\nabla_\mu e^{\ph b}_{\nu}=
e^{\ph c}_{\mu}\,\gamma^{ba}_{\ph\ph c}\;,
\end{equation}
where the $\omega_{\mu}^{\ph ab}$'s denote the so-called \emph{spin connections} and $\gamma_{abc}=e^{\mu}_{\ph c} e^{\nu}_{\ph b} \nabla_\mu e_{\nu a}$ are the Ricci Rotation Coefficients ($\nabla_\mu$ is the usual coordinate covariant derivative). Furthermore, spin connections are able to restore the correct Dirac algebra in curved space-time, \emph{i.e.}, $D^{\sss}_\mu\;\gamma^{\nu}=0$ \cite{hammond}.

This picture suggests, in appearance, the description of gravity as a gauge model \cite{cho1}. In fact, the curvature tensor and the Einstein-Hilbert Action write as:
\begin{equation}\label{Riemann}
R^{\ph\ph ab}_{\mu\nu}=\p_\nu\omega_{\mu}^{\ph ab}-
\p_\mu\omega_{\nu}^{\ph ab}+\mt{F}^{ab}_{cdef}\omega_{\mu}^{\ph cd}\omega_{\nu}^{\ph ef}\;,
\q
S_{EH}(e,\omega)=-\tfrac{1}{4}{\textstyle \int}
\mathrm{det}(e)\,d^{4}x\;\;e^{\ph\mu}_{a}e^{\ph\nu}_{b} R^{\ph\ph ab}_{\mu\nu}\;.
\end{equation}
and variation of $S_{EH}$ wrt connections leads to the II Cartan Structure Equation
\begin{equation} \label{Cartan eq}
\p_\mu e^{\ph a}_{\nu} -\p_\nu e^{\ph a}_{\mu}-\omega_{\mu}^{\ph ab}e_{\nu b}
+\omega_{\nu}^{\ph ab}e_{\mu b}=0\;.
\end{equation}
In the standard approach, denoting with $\Lambda_a^{b}$ the Lorentz matrix, spin connections transform like gauge vectors under infinitesimal local Lorentz transformations $\Lambda_{a}^{b}=\delta_{a}^{b}+\epsilon_{a}^{b}$, \emph{i.e.},
\begin{equation}\label{gaugetr}
\omega_{\mu}^{\ph ab}\stackrel{L}{\to} \omega_{\mu}^{\ph ab}-\p_\mu\epsilon^{ab}+
\tfrac{1}{4}\mt{F}^{ab}_{cdef}\epsilon^{cd}\omega_{\nu}^{\ph ef}\;.
\end{equation}
Furthermore, we underline that the $\omega_{\mu}^{\ph ab}$'s behave like ordinary vectors under general coordinate transformations (\emph{i.e.}, world transformations). Here, the presence of tetrad fields (introduced by the Principle of General Covariance) is an ambiguous element for the gauge paradigm. In fact, spin connections can be uniquely determined as functions of tetrads (in terms of the Ricci Rotation Coefficients $\gamma_{abc}$) and this relation generates an ambiguity in the interpretation of the $\omega_{\mu}^{\ph ab}$'s as the only fundamental fields of the gauge scheme since the theory were based on two dependent degrees of freedom.

\section{A Novel Approach for a Gauge Theory of the LG}In the model we propose here, the \emph{key point} is that, if we are able to show (as we will demonstrate in the following) how diffeomorphisms can induce local Lorentz transformations, we can conclude that the $\omega_{\mu}^{\ph ab}$'s can no longer be regarded as gauge potentials for the LG, because of their transformation properties: do they behave like gauge fields or ordinary (coordinate) vectors? In this sense, we hypothesize that new gauge fields must be added to restore the Lorentz invariance of the theory. By other words, a strangeness arises as far as spinor fields are analyzed under diffeomorphism-induced Lorentz rotation: fermions are expected to be coordinate scalars and to transform according the usual laws under Lorentz rotations. They live in the tangent bundle without experiencing coordinate changes and, if the two transformations formally overlap, a puzzle on the nature of spinors comes out \cite{mpla}. 

The fundamental hypothesis of our model is in assuming that \emph{spin connections transform like vectors if diffeomorphism-induced rotations are implemented}. In this scheme, new gauge fields $A_\mu^{\ph ab}$, transforming according to their Lorentz indices, are the only fields able to restore Lorentz invariance when local rotations are induced by coordinate changes. In fact, the nature of gauge potentials is naturally lost by spin connections $\omega_\mu^{\ph ab}$, which are assumed to behave like tensors only. 

Let us now demonstrate that the correspondence between coordinate transformations and local rotations takes place only if we deal with infinitesimal \emph{isometric diffeomorphisms}:
\begin{equation}\label{newdiff}
x^{\mu}\to x^{\primo\mu}=x^{\mu}+\xi^{\mu}(x)\;,\qq
\qquad\;\;\nabla_{\mu}\xi_{\nu}+\nabla_{\nu}\xi_{\mu}=0\;,
\end{equation}
and the following transformation of the basis vectors is induced
\begin{equation}\label{taylor}
e_{\mu}^{\ph a}(x)\stackrel{D}{\to}e_{\mu}^{\primo\ph a}(x^{\primo})=
e_{\nu}^{\ph a}(x)\;\nicefrac{\p x^{\nu}}{\p x^{\primo\mu}}=
e^{\ph a}_{\mu}(x)-e^{\ph a}_{\nu}(x)\;\nicefrac{\p \xi^{\nu}}{\p x^{\primo\mu}}\;.
\end{equation}
If we deal with an infinitesimal local Lorentz transformation $\Lambda_{a}^{b}(x)=\delta^b_a +\epsilon^{b}_{a}(x)$, we get, up to the leading order, the tetrad change (evaluated in $x^{\primo}$ of eq. \reff{newdiff}):
\begin{equation}
e_{\mu}^{\ph a}(x)\stackrel{L}{\to}e_{\mu}^{{\scriptscriptstyle \prime}\ph a}(x^{{\scriptscriptstyle \prime}})=\Lambda_{a}^{b}(x^{\scriptscriptstyle \prime})e_{\mu}^{\ph a}(x^{{\scriptscriptstyle \prime}})=
e_{\mu}^{\ph a}(x^{{\scriptscriptstyle \prime}})+e_{\mu}^{\ph b}(x)\,\epsilon^{a}_{b}(x)\;.
\end{equation}
We can infer that the two transformation laws formally overlap if we assume the condition: 
\begin{equation}\label{rotation-parameter}
\epsilon_{ab}=\nabla_{[a}\xi_{b]}-\gamma_{abc}\,\xi^c\;,
\end{equation}
where, to pick up local Lorentz transformations from the set of generic diffeomorphisms, the isometry condition $\nabla_{(\mu}\xi_{\nu)}=0$ has to be taken into account in order to get the antisymmetry condition $\epsilon_{ab}=-\epsilon_{ba}$ for the infinitesimal parameter $\epsilon_{ab}$.

\section{Spinorial Lie Derivative} It can be interesting to compare our approach with the one of \cite{ortin, kosmann}, where the formalism of the \emph{Lie Derivative} is extended to fermion fields. Spinors can only be introduced in the tangent bundle of curved space-time by using the well-known Weyl formalism \cite{weyl}. Such scheme makes use of tetrads (described in Sec. 2) constituting an orthonormal basis in tangent space which is covariant wrt local transformations preserveing the orthonormality, \emph{i.e.}, local Lorentz $SO(3,1)$ transformations. As discussed above, spinors behave as scalars under diffeomorphisms and, in the Weyl approach working in curvilinear coordinates on a flat Minkowski background, one can find formally that Lorentz transformations coincide with diffeomorphisms which do not affect the spinorial indices. Thus, also using Weyl formalism, the nature of the spinor transformations becomes ambiguous. 

Without using a gauge approach, such a puzzle was treated by Y. Kosmann \cite{kosmann} (see also \cite{ortin} and references therein) defining a \emph{spinorial Lie derivative} $\mathfrak{L}_{\xi}$ wrt Killing vector fields $\xi^\mu$ of eq. \reff{newdiff} (the standard Lie derivative reads $\pounds_{\xi}\psi =\xi^{\mu}\partial_{\mu}\psi$):
\begin{equation}\label{eq:LLonspinors}
\mathfrak{L}_{\xi}\;\psi \equiv \xi^{\mu}D^{\sss}_\mu\psi + 
\tfrac{1}{4}\;D^{\sss}_{[a}\xi^{\phantom{\sss}}_{b]}\Sigma^{ab}\,\psi\;.  
\end{equation}
Such a derivative, which is Lorentz-covariant, does not enter the Dirac Lagrangian as in a gauge model, but the second term of the rhs of eq. \reff{eq:LLonspinors} states the action of the isometric diffeomorphisms on spinor fields as an infinitesimal Lorentz rotation described the parameter $\epsilon_{ab}=\nicefrac{1}{2}\;D^{\sss}_{[a}\xi^{\phantom{\sss}}_{b]}$ (see eq. \reff{rotation-parameter}). 

In \cite{ortin}, an interesting example is addressed: in a Minkowski space-time $g_{\mu\nu}=\eta_{\mu\nu}$, choosing the gauge $e_\mu^{\ph a}=\delta_\mu^{\ph a}$, one can suppose the infinitesimal diffeomorphism described by the Killing vector $\xi^\mu=\theta^{\mu}_{\nu}x^{\nu}$ (where  $\theta^{\mu\nu}=-\theta^{\nu\mu}=const.$). As already discussed, this is formally a standard infinitesimal Lorentz rotation and one gets
\begin{equation}
\mathfrak{L}_{\xi}\;\psi=\xi^{\mu}\partial_{\mu}\psi 
+\tfrac{1}{4}\;\theta_{ab}\Sigma^{ab}\psi\;,
\end{equation}
as expected using the standard spinor formalism. This way, $\mathfrak{L}_{\xi}\psi$ represents a Lorentz covariantization of the standard Lie derivative $\pounds_{\xi}\psi =\xi^{\mu}\partial_{\mu}\psi$ generating a Lorentz rotation which trivializes the holonomy.

\section{Spinors and Gauge Theory of the LG in Flat Space-Time} In such a scheme, the usual spin connections $\omega_\mu^{\ph ab}$ can be set to zero choosing the gauge $e_\mu^{\ph a}=\delta_\mu^{\ph a}$ (in general, they are allowed to be non-vanishing) and spin$-\nicefrac{1}{2}$ fields are described by the Lagrangian density
\begin{equation}
\mathcal{L}_F=\tfrac{i}{2}\;\bar{\psi}\gamma^ae^{\mu}_{\ph a}\p_{\mu}\psi-
\tfrac{i}{2}\;e^{\mu}_{\ph a}\p_{\mu}\bar{\psi}\gamma^a\psi
\end{equation}\newcommand{\as}{\scriptscriptstyle{(A)}}and they have to recognize the isometric components of the diffeomorphism as a local Lorentz transformation since a spinor can noway be a Lorentz scalar. In this scheme, new Lorentz connections $A_\mu^{\ph ab}$ have to be introduced for matter fields since, for assumption, the standard connections $\omega_{\mu}^{\ph ab}$ do not follow Lorentz gauge transformations and they are not able to restore Lorentz invariance. An infinitesimal local Lorentz transformation \reff{LG} act on the spinor as $\psi(x)\to S\;\psi(x)$ and $\gamma$ matrices transform like Lorentz vectors, \emph{i.e.},
$S\,\gamma^{{a}}\,S^{-1}=(\Lambda^{-1})^{{a}}_{{b}}\,\gamma^{{b}}$. The gauge invariance is restored by the covariant derivative
\begin{equation}
D^{\as}_\mu\psi=(\p_\mu-\tfrac{i}{4}\,A_\mu^{\ph ab}\,\Sigma_{ab})\,\psi\;,\qq
A_\mu^{\ph ab}\rightarrow A_\mu^{\ph ab}-\p_\mu\epsilon^{ab}+4\mathcal{F}^{\,ab}_{cdef}\,\, \epsilon^{ef}\,A^{\ph cd}_\mu\;,
\end{equation}
where $A_\mu^{\ph ab}\neq\omega_\mu^{\ph ab}$ represent natural Yang-Mill fields associated to the LG, living in the tangent bundle. A Lagrangian $\mt{L}_A$ associated to the gauge connections can be constructed by the introduction of the gauge field strength $F_{\mu\nu}^{\ph\ph ab}$, \emph{i.e.},
\begin{equation}\label{action-for-A}
F_{\mu\nu}^{\ph\ph ab}=\p_\mu A_\nu^{\ph ab}-\p_\nu A_\mu^{\ph ab}
+\tfrac{1}{4}\mt{F}^{ab}_{cdef}A_\mu^{\ph cd}A_\nu^{\ph ef}\;,\qq
\mt{L}_A=-\tfrac{1}{4}\;F_{\mu\nu}^{\ph\ph ab}F^{\mu\nu}_{\ph\ph ab}\;.
\end{equation}
In flat space, the only real dynamical fields are the Lorentz gauge fields.

\section{Curved Space-Time and the Role of Torsion} In what follows, we provide, in the First-Order Approach \cite{afldb}, a link between the dynamics of the contortion field $\mt{K}_\mu^{\ph ab}$ and the new Lorentz gauge connections $A_\mu^{\ph ab}$. The need to introduce the $A_\mu^{\ph ab}$ fields in curved space-time is motivated by the restoration of the local invariance under diffeomorphism-induced Lorentz transformations, while spin connections $\omega_\mu^{\ph ab}$ allow one to recover the proper Dirac algebra. 

Considering the Riemann-Cartan space $U^{4}$ filled with \emph{general} affine connections $\tilde{\Gamma}_{\mu\nu}^{\rho}$, the torsion field $\mt{T}_{\mu\nu}^{\rho}$ is defined as
$\mt{T}_{\mu\nu}^{\rho}=\tilde{\Gamma}_{[\mu\nu]}^{\rho}$ and the II Cartan Structure Equation writes \cite{kim}
\begin{equation}\label{Cartan eq general}
\p_\mu e^{\ph a}_{\nu} -\p_\nu e^{\ph a}_{\mu}-\tilde{\omega}_{\mu}^{\ph ab}e_{\nu b}
+\tilde{\omega}_{\nu}^{\ph ab}e_{\mu b}=
e^{\ph a}_{\rho}\,\tilde{\Gamma}_{[\mu\nu]}^{\rho}=
e^{\ph a}_{\rho}\mt{T}_{\mu\nu}^{\rho}=
\mt{T}_{\mu\nu}^{\ph\ph a}\;.
\end{equation}
The total connections $\tilde{\omega}_{\mu}^{\ph ab}$, solution of this equation, are
\begin{equation}\label{connection}
\tilde{\omega}_{\mu}^{\ph ab}=\omega_{\mu}^{\ph ab}+\mt{K}_{\mu}^{\ph ab}\;,
\end{equation} 
here $\mt{K}_{\mu}^{\ph ab}$ is the contortion field derived by the usual relation 
$\mt{K}^{\mu}_{\nu\rho}=-\tfrac{1}{2}(\mathcal{T}^{\mu}_{\nu\rho}-\mathcal{T}_{\rho\nu}^{\mu}+\mathcal{T}^{\mu}_{\nu\rho})$.

To establish the proper geometrical interpretation of the new gauge fields $A_\mu^{\ph ab}$, let us now introduce generalized connections $\bar{\omega}_\mu^{\ph ab}$ and postulate the following interaction term 
\begin{equation}\label{interacting term}
S_{conn}=2{\textstyle \int} \mathrm{det}(e)\,d^{4}x\;\;
e_{\ph a}^{\mu}e_{\ph b}^{\nu}\;\bar{\omega}_{\mu c}^{\ph [a}\,A_{\nu}^{\ph bc]}\;.
\end{equation}
In such an approach, the action describing the dynamics of the fields $A_\mu^{\ph ab}$ can be usually derived form the gauge Lagrangian $\mathcal{L}_A$ \reff{action-for-A}, while the action that accounts for the generalized connections can be taken as the action $S_{EH}$ \reff{Riemann}, but now the projected Riemann tensor $R_{\mu\nu}^{\ph\ph ab}$ in \reff{Riemann} is constructed by the $\bar{\omega}_\mu^{\ph ab}$'s. Collecting all terms together, one can get the total action for the model. If fermion matter is absent, variation of the total action wrt connections $\bar{\omega}_\mu^{\ph ab}$ gives the generalized equation
\begin{align}\label{equation for omega}
\p_\mu e^{\ph a}_{\nu} -\p_\nu e^{\ph a}_{\mu}-\bar{\omega}_{\mu}^{\ph ab}e_{\nu b}+
\bar{\omega}_{\nu}^{\ph ab}e_{\mu b}=
A_{\mu}^{\ph ab}e_{\nu b}-A_{\nu}^{\ph ab}e_{\mu b}\,,
\end{align}
which admits the solution
\begin{equation}\label{solution vacuum}
\bar{\omega}_\mu^{\ph ab}=\omega_\mu^{\ph ab}+A_{\nu}^{\ph ab}\;,
\end{equation}

As a result, comparing the expression above with the solution (\ref{connection}), we show how the new gauge fields $A_\mu^{\ph ab}$ mimic, \emph{on shell}, the dynamics of the contortion field $\mt{K}_\mu^{\ph ab}$. To conclude, we propose that contortion represents that field which is responsible for a correct Lorentz covariantization of the spinor dynamics under isometric diffeomorphism-induced local rotations of the tetrad basis.

\section{Concluding Remarks} In the proposed model, the key point has been fixing the equivalence between isometric diffeomorphisms and local Lorentz transformations. In fact, under the action of the former, spin connections behave like a tensor and are not able to ensure invariance under the corresponding induced local rotations. This picture has led us to infer the existence of (metric-independent) compensating fields of the LG which interact with spinors. In curved space-time, a mathematical relation between the Lorentz gauge fields and contortion has been found from the II Cartan Structure Equation if a (unique) interaction term between the gauge fields and generalized internal connections is introduced.


\begin{thebibliography}{0}
\bibitem{hammond}
R.T. Hammond, {\it Rep. Prog. Phys.} {\bf 65}, 599 (2002).
 
\bibitem{cho1}
Y.M. Cho, {\it Phys. Rev. D} {\bf 14}, 2521 (1976).

\bibitem{mpla}
N. Carlevaro, O.M. Lecian and G. Montani, {\it Mod. Phys. Lett. A} {\bf 24}, 415 (2009).

\bibitem{ortin}
T. Ort\'in, {\it Class. Quant. Grav.} {\bf 19}, L143 (2002).

\bibitem{kosmann}
Y. Kosmann, {\it Ann. Mat. Pura Appl. (IV)} {\bf 91}, 317 (1972).

\bibitem{weyl}
L. O'Raifeartaigh, {\it The Dawning of Gauge Theory} (Princeton Univ. Press, 1997). 

\bibitem{afldb}
N. Carlevaro, O.M. Lecian and G. Montani, {\it Ann. Fond. L. de Broglie} {\bf 32}, 281 (2007).

\bibitem{kim}
S-W. Kim and D.G. Pak, {\it Class. Quant. Grav.} {\bf 25}, 065011 (2008).
  
\end{thebibliography}
\end{document}